# IMPROVING PERFORMANCE OF ENGLISH-HINDI CROSS LANGUAGE INFORMATION RETRIEVAL USING TRANSLITERATION OF QUERY TERMS

Saurabh Varshney[1] and Jyoti Bajpai[2]

[1]Department of Computer Engineering, GLA University, Mathura, India
[2] Department of Computer Engineering, GLA University, Mathura, India

***ABSTRACT***

*The main issue in Cross Language Information Retrieval (CLIR) is the poor performance of retrieval in terms of average precision when compared to monolingual retrieval performance. The main reasons behind poor performance of CLIR are mismatching of query terms, lexical ambiguity and un-translated query terms. The existing problems of CLIR are needed to be addressed in order to increase the performance of the CLIR system. In this paper, we are putting our effort to solve the given problem by proposed an algorithm for improving the performance of English-Hindi CLIR system. We used all possible combination of Hindi translated query using transliteration of English query terms and choosing the best query among them for retrieval of documents. The experiment is performed on FIRE 2010 (Forum of Information Retrieval Evaluation) datasets. The experimental result show that the proposed approach gives better performance of English-Hindi CLIR system and also helps in overcoming existing problems and outperforms the existing English-Hindi CLIR system in terms of average precision.*

## 1. INTRODUCTION

We are rapidly constructing the broad network architecture for transferring information across national barriers, but much remains to be done before linguistic boundaries can be better as effectively as geographic ones [1]. Now a days, peoples have more likely to interest on global things like education, economy, business, marketing, research etc. because of that peoples are interested to collect information and data of other regions of the world. The one and only medium for doing this, is the Internet. But we also knows users are more likely to retrieve information in a language in which a user is more comfortable or we can say that user wants information in his/her native language to understand the language of documents more easily. Accessing information in a host language is clearly important for many users. In India, about 70% of peoples know Hindi as a primary language while based on human development survey in 2012; there are only 10.35 % peoples in India who are the English speakers. India is third country that has largest number of internet users but when we talk about penetration means total population, in India only 12.6% of people are the internet user which decrease the rank of India on to 164th position based on survey. And we also know that entering query in another language to retrieve documents is very difficult to the user. So, the conclusion is that, there should be require a tool that takes query in English language and provides relevant information in our native language.

The Internet environment gives the benefits for this issue by providing Cross Language Information Retrieval (CLIR) technology. Because of big bang of on-line non-English webpage's, CLIR systems have become progressively more important in recent years [2]. CLIR filling the gap of linguistic barrier by allow a user to search in one language and retrieve documents in another language.

CLIR is important because of various reasons that are as follows:

- Sometime, we are not able to find an appropriate query to find top relevant document. Like if I want to download Ramcharitramaanas in Hindi language. If I enters query in Hindi language (like रामचरित्रमानस) than it gives more promising result as compared to the English query (like Ramcharitramaanas) because sometimes documents are completely in a single language(like Hindi) because of that user query based IR system cannot retrieve such documents.
- CLIR increases the percentage of users in internet because it provides the information in their native language.

But we also know in India, there are many words that are known because of their English meaning like computer, cricket, bank and many more, people's do not knows their Hindi meaning and even sometime peoples prefer English words to makes sentences. We also know very well that information retrieval models works on similarity between query and documents. After the query translation in English-Hindi CLIR, if we get a Hindi meaning of such types of words then definitely the performance of CLIR system will decrease because of mismatching between query terms and documents.

## 2. Related Work

Considerable amount of work is already done in English-Hindi CLIR. The different-different approaches for retrieving information from CLIR system have some advantages and disadvantages. Lisa, et al. [3] in 1998 proposed a method for resolving ambiguity in query translation and phrasal translation by using statistics co-occurrence analysis from unlinked corpus and combines this technique with other techniques for resolving ambiguity and achieve more than 90% of CLIR performance while compared to the monolingual performance and also author compared their method with machine translation and parallel corpus techniques and they proved that good performance of retrieval can be achieved without the use of complex resources. Kyung-Soon et al. [4] in 2002 proposed a method to implicitly resolve ambiguities in Korean-English CLIR system using dynamic incremental clustering approach means the clusters are incrementally created for the top ranked documents for a particular query and next time when the same query will fired than the weight of each retrieved document is recalculated by using these clusters. Dong Zhou et al. [5] in 2008 developed a disambiguation strategy for determining the correct translation for a given query by using novel graph based analysis of co-occurrence information and also developed a new approach to translate OOV (Out Of Vocabulary) terms means the words that are commonly not found in dictionary like, proper names, location, address etc. Sujoy Das et al. [6] in 2010 investigated the influence of query expansion using WordNet in English-Hindi CLIR system. Author used shabdanjali dictionary for English-Hindi query translation and expands Hindi queries by using Hindi WordNet and used nine different strategies for query expansion. Based on the results, author observed that query expansion using Hindi WordNet is not more effective and not gives a better performance while compared to monolingual performance. S.M. Chaware et al. [7] in 2011 proposed an approach to build ontology from relational database with the help of some additional rules that can also be used for cross lingual information retrieval. The ontology approach is based on user requirements that give overall knowledge of domain to the user.

## 3. PROPOSED METHODOLOGY

The system which takes user query in one language and retrieves relevant documents in other language is known as cross language information retrieval system. Studies say that the performance of CLIR is still poor as compared to Mono-lingual performance and also the problem of ambiguity in query translation down the performance of CLIR in term of recall and precision. Several methods have been already proposed in order to solve the given problem of CLIR like query expansion, co-occurrence statistics, Clustering etc, but still the performance of English-Hindi CLIR is not as good as compared to monolingual IR performance. The most common reasons behind the poor performance of CLIR are as follows:

- Lack of availability of resources like Bilingual dictionary, mismatching of out of vocabulary (OOV) terms, stemmer, part of speech (POS) tagger etc.
- Multiple representations of query words (Lexical ambiguity).
- Problem in encoding the text (UTF-8)
- Poor matching and translation techniques.

These problems are due to the limitations in the existing approaches. Therefore, the limitations of the existing approaches need to be further inquired towards achieving the increase in the performance English-Hindi CLIR. The main aim for inquiring the limitations of existing approaches and to develop a new approach to find out all the relevant information from CLIR with higher and higher recall and with no or very less amount of irrelevant information retrieved according to the query given by the user. So, in our approach, we used transliteration of each query terms to make all possible combination of query.

The proposed algorithm for English-Hindi CLIR is given below that shows the step by step process of English-Hindi CLIR.

1. User enters the query in English language $Q_E$.
2. Finds all terms from $Q_E$ and translate those terms into Hindi language using English-Hindi dictionary and naming them as $\{t_1, t_2, t_3 \ldots\ldots\ldots t_K\}$.
3. Finds all terms from $Q_E$ and transliteration those terms into Hindi language using Itrans tool and naming them as $\{t'_1, t'_2, t'_3 \ldots\ldots t'_K\}$
4. Mapping terms $\{t_1, t_2, t_3 \ldots\ldots\ldots t_K\} = \{t'_1, t'_2, t'_3 \ldots\ldots\ldots t'_K\}$
5. Translate English query $Q_E$ into Hindi query $Q_H$.
6. Making all the possible combination of Hindi Query $Q_H$ using $\{t'_1, t'_2, t'_3 \ldots\ldots\ldots t'_K\}$ without replacement of term position up to $2^k$ times, where k is the number of terms in $Q_E$.
7. Calculate the mean average precision (MAP) of all possible queries that is generated from step 6 and from them choose the best query and named that query as $Q'_H$.
8. All the relevant documents generate by the query $Q'_H$ gives to the user.

To understand the algorithm, we consider an example that is shown below:

1. User enters the query $Q_E$ = {Democracy in India}.
2. Finds all the terms from $Q_E$ i.e. {Democracy, India} and translate them into Hindi language as {लोकतंत्र, भारत} and naming them as $\{t_1, t_2\}$.
3. Finds all the terms from $Q_E$ and transliteration those terms into Hindi language using Itrans tool as {डेमोक्रेसी, इंडिया} and naming them as $\{t'_1, t'_2\}$.
4. Mapping terms $\{t_1, t_2, t_3 \ldots\ldots\ldots t_K\} = \{t'_1, t'_2, t'_3 \ldots\ldots\ldots t'_K\}$ means

$t_1 = t'_1$ (लोकतंत्र = डेमोक्रेसी) and $t_2 = t'_2$ (भारत = इंडिया)

5. Translate English query $Q_E$ into Hindi query$Q_H$.

   $Q_H$ = {भारत में लोकतंत्र}

6. Making all the possible combination of Hindi Query $Q_H$ using { $t'_1$, $t'_2$} without replacement of term position up to $2^k$ times means $2^2$ times

   $Q_H$ = (भारत में लोकतंत्र | इंडिया में लोकतंत्र | भारत में डेमोक्रेसी | इंडिया में डेमोक्रेसी )

7. Calculate the mean average precision (MAP) of all possible queries that is generated from step 6 and from them choose the best query and named that query as $Q'_H$.
8. All the relevant documents generate by the query $Q'_H$ gives to the user.

### 3.1. Query Translation

Translation of query from one language to other language is known as query translation. Query translation is a crucial step in CLIR system because all problems come from this step like mismatching of query terms, ambiguities, poor retrieval performance etc. There are various approaches are used to translate user query like Bilingual dictionary, parallel corpus, online translator etc. In this paper, we have used 'Shabdanjali' Multi-lingual Readable Dictionary as a lexicon resource for translating English to Hindi query. The dictionary was developed in IIIT Hyderabad. This dictionary is available in ISCII conversion. So, a conversion from ISCII to UTF-8 encoding code is required. The other inbuilt tools/resources that help to translate English query to Hindi query is shown in table 1.

TABLE 1 Tools used for query translation

| **Resources** | **Tool Used** |
|---|---|
| **Morphological Analysis** | ittoolbox |
| **POS tagger** | Stanfort POS tagger |
| **Transliteration** | I-Trans |
| **STOP word** | List of 480 Stop words |
| **Stemmer** | Porter Stemmer |

## 4. EXPERIMENTS

Experiment is performed on FIRE (Forum of Information Retrieval Evaluation) 2010 datasets. FIRE 2010 datasets consists of set of user queries in terms of 'Title' field, 'Description' field and 'narrative' field, set of documents and qrel files which gives a list of relevant documents for a queries. The experiment performed for English-Hindi CLIR to retrieve Hindi documents using English queries. In this paper we used only 'Title' field of queries. Test data collection describe in table 2 as follows:

TABLE 2 Statistics of FIRE 2010 data collection

| **Metrics** | **CLIR** |
|---|---|
| **Query Language** | English |
| **Document Language** | Hindi |
| **No. of Queries** | 50 |
| **No. of Documents** | 1,49,481 |
| **Size of Documents** | 1.36 GB |
| **Avg. no of rel. documents per** | 18 |
| **query** | |

We obtain results of two approaches, first EHT approach which means English-Hindi CLIR using title field of query and second is EHRT approach which means English-Hindi CLIR using refined title field of query using our approach. The results of both approaches are shown in term of interpolation recall-precision average in table 3.

Table 3 Recall-Precision Average

| **Metric** | **EHT** | **EHRT** |
|---|---|---|
| **No. of query** | 50 | 50 |
| **No. of retrieve documents** | 50,000 | 50,000 |
| **No. of rel. documents** | 915 | 915 |
| **No. of retrieve relevant** | 859 | 880 |
| **At 0.00** | 0.6707 | 0.7221 |
| **At 0.10** | 0.5852 | 0.6233 |
| **At 0.20** | 0.5088 | 0.5421 |
| **At 0.30** | 0.4395 | 0.4741 |
| **At 0.40** | 0.3827 | 0.4202 |
| **At 0.50** | 0.3334 | 0.3697 |
| **At 0.60** | 0.2935 | 0.3260 |
| **At 0.70** | 0.2478 | 0.2737 |
| **At 0.80** | 0.1929 | 0.2150 |
| **At 0.90** | 0.1364 | 0.1517 |
| **At 1.00** | 0.1074 | 0.1153 |
| **MAP** | **0.3324** | **0.3609** |
| **At 5 docs** | 0.4240 | 0.4360 |
| **At 10 docs** | 0.3800 | 0.4060 |
| **At 15 docs** | 0.3373 | 0.3667 |
| **At 20 docs** | 0.3000 | 0.3260 |
| **At 30 docs** | 0.2460 | 0.2607 |
| **At 100 docs** | 0.1174 | 0.1236 |
| **At 200 docs** | 0.0672 | 0.0708 |
| **At 500 docs** | 0.0325 | 0.0334 |
| **At 1000 docs** | 0.0171 | 0.0176 |
| **R-precision** | **0.3260** | **0.3459** |

## 4.1. Performance graph

From the table 2, it is clear that our English-Hindi CLIR gives better performance as compared to original title field queries of English-Hindi CLIR in terms of MAP (mean average precision). Figure 1 shows the interpolation recall-precision average of both systems.

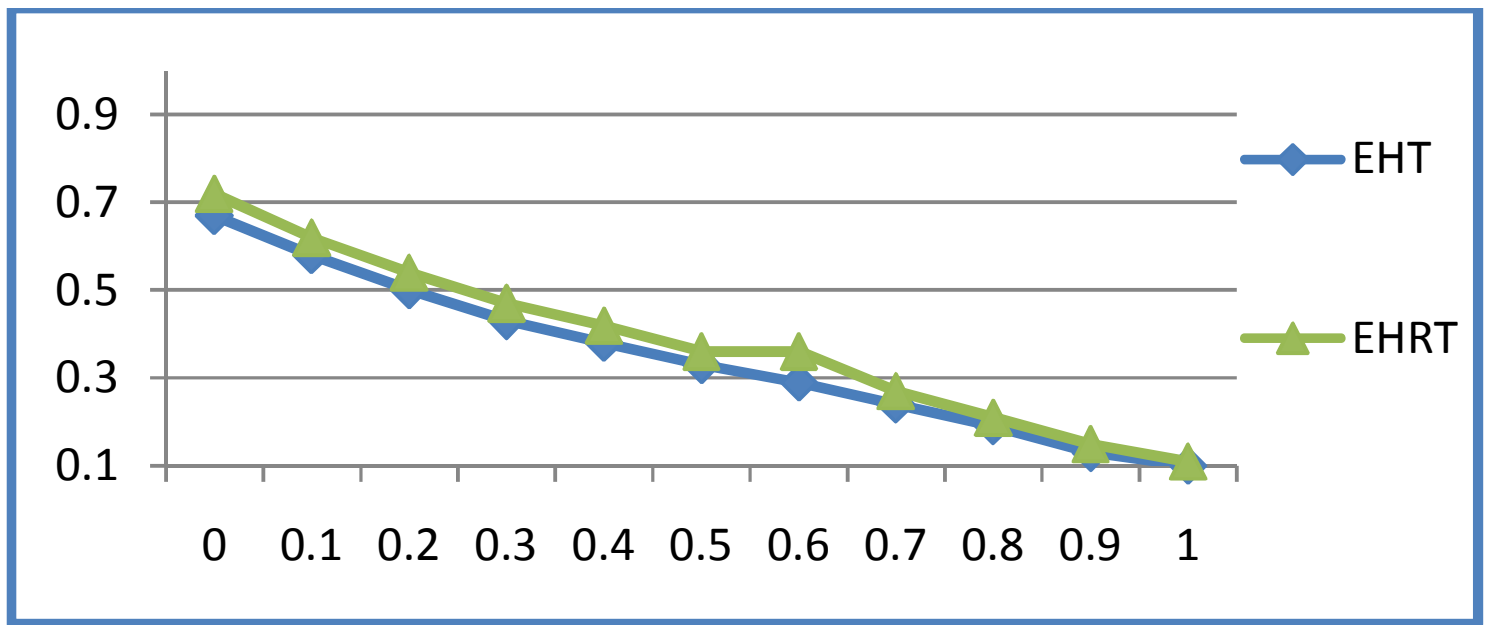

Figure 1 Interpolation Recall-Precision Averages

## 5. Conclusion

The hurdle problem in CLIR is poor performance when compared to monolingual IR performance because of query term mismatching, un-translated query words, multiple representations of query terms etc. There are considerable amount of work is already done in English-Hindi CLIR. The different-different approaches for retrieving information from CLIR system are discussed in related work part. Those earlier proposed approaches have some advantages and disadvantages. In order to make these approaches to be more efficient and effective practically the limitations of those approaches need to be further inquired towards achieving the increase in the performance of the English-Hindi CLIR system.

In this paper, we used transliteration of each query terms of English query to make all possible combination of Hindi query without replacement of query term position in order to improving of English Hindi CLIR. The average mean precision of EHT and EHRT strategies are .3324 and .3721. The experimental results show that the proposed approach for refined 'title' field of queries gives more relevant information as compared to original 'title' field of FIRE 2010 queries. Therefore proposed approach can helps to improving the performance of English-Hindi CLIR and retrieving rich and high quality of information with increase in average precision. This approach is language independent means this approach works on any combination of languages.

## Author

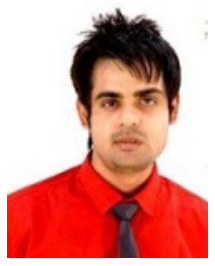

**Saurabh Varshney** obtained his bachelor's degree from UPTU University, Lucknow, India. Now he is currently an M.Tech student under the supervision of Asstt. Prof. Jyoti Bajpai. His research is centred on performance of English-Hindi Cross Language Information Retrieval System.